\documentclass[b5paper,twoside]{jpconf}  
\usepackage{graphicx}

\begin{document}

\title[Statistical Tests of Cepheid P-L Relations]{Statistical Tests for the Metallicity Dependency of the Synthetic Cepheid Period-Luminosity Relations in IRAC Bands}  

\author[Ngeow et al.]{Chow-Choong Ngeow$^1$, Marcella Marconi$^2$, Ilaria Musella$^2$, Michele Cignoni$^3$ and Shashi Kanbur$^4$}

\address{$^1$Graduate Institute of Astronomy, National Central University, Jhongli City, 32001, Taiwan}
\address{$^2$Osservatorio Astronomico di Capodimonte, Via Moiariello 16, 80131 Napoli, Italy}
\address{$^3$Department of Astronomy, Bologna University, via Ranzani 1, 40127 Bologna, Italy}
\address{$^4$Department of Physics, State University of New York at Oswego, Oswego, NY 13126, USA}

\ead{cngeow@astro.ncu.edu.tw} 

\begin{abstract}

The mid-infrared (MIR) period-luminosity (P-L) relations for Cepheids will be important in the {\it JWST} era, as it holds the promise of deriving the Hubble constant within 2\% accuracy. It is expected that the MIR P-L to be insensitive to metallicity. In this work, we test this assumption of metallicity independent of the IRAC band P-L relation by applying well-known statistical methods to the synthetic P-L slopes from a series of pulsating models with known metallicity. The statistical tests suggest that the P-L slopes in MIR are linearly depending on metallicity.

\end{abstract}

\section{Introduction}

The Cepheid mid-infrared (MIR) period-luminosity (P-L, also known as Leavitt Law) relations will be important in the {\it James Webb Space Telescope (JWST)} era, as it holds the promise of deriving the Hubble constant within 2\% accuracy \cite{fre10}. This is because one of the systematic error in Hubble constant -- the extinction is negligible in MIR. Also, the MIR P-L relation is expected to be insensitive to metallicity. In this work, we test the assumption of metallicity independent of the IRAC band P-L relations: we apply statistical tests to test the correlations between the slopes of the synthetic P-L relation and metallicity.

\section{Models and Statistical Tests}

\begin{figure}[h]
\begin{minipage}{14pc}
\includegraphics[width=14pc]{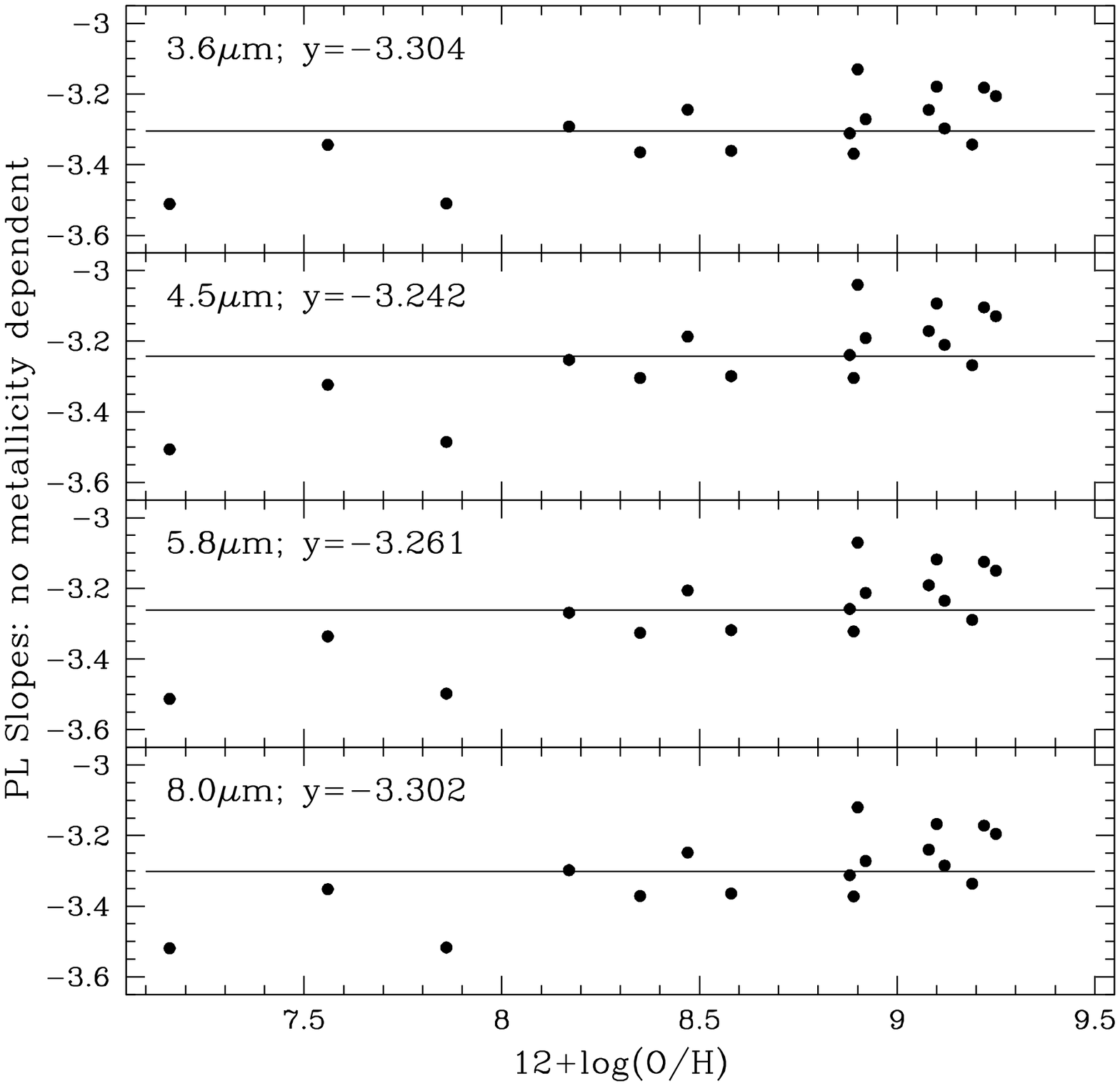}
\caption{\label{label1}Fitting of the constant regressions to the IRAC band P-L slopes, assuming the slopes are independent of metallicity.}
\end{minipage}\hspace{2pc}%
\begin{minipage}{14pc}
\includegraphics[width=14pc]{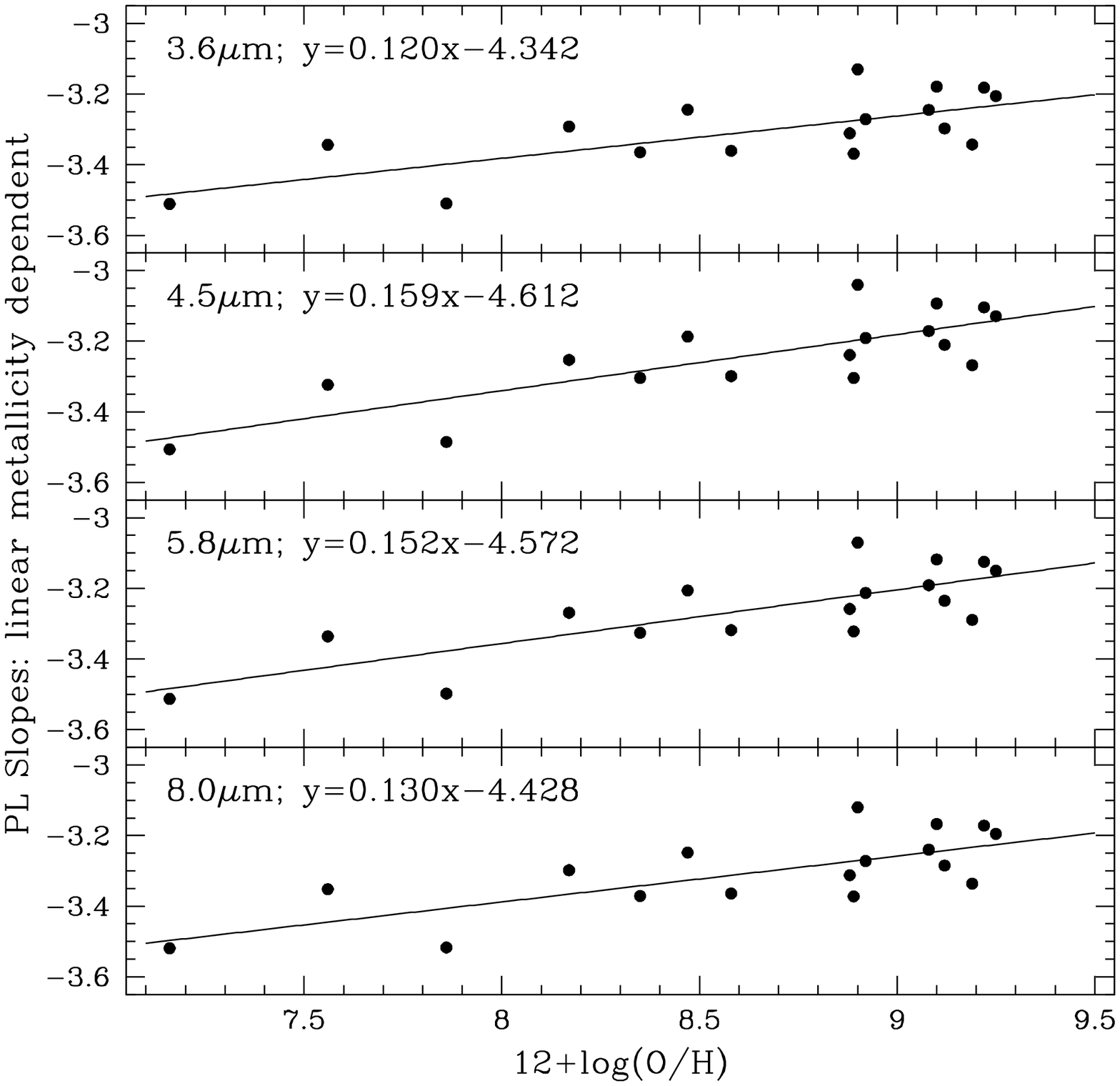}
\caption{\label{label2}Fitting of the linear regressions to the IRAC band P-L slopes, assuming the slopes are linearly depending on metallicity.}
\end{minipage}
\end{figure}

Details of deriving the synthetic P-L relations from pulsation models are given in \cite{nge11}. The metallicity, $12 + \log(O/H)$, for each sets of IRAC band PL relations is known. We test the following relations between the metallicity and the synthetic IRAC band P-L slopes: {\bf Case (I)} The P-L slopes are independent of metallicity; and {\bf Case (II)} The P-L slopes are linearly depending on metallicity. Figure \ref{label1} and \ref{label2} show the fitting of these two cases to the synthetic P-L slopes, respectively.  

We have applied the Akaike's Information Criterion ($AIC$) and Schwarz's Bayesian Criterion ($SBC$), methods that widely used in statistical literature for models selection, to test the metallicity dependency of synthetic IRAC band P-L slopes. Model with smallest AIC and SBC will be the preferred model. The results from $AIC$ test are as follow. $3.6\mu\mathrm{m}$: $AIC(I)=-74.6$, $AIC(II)=-85.3$; $4.5\mu\mathrm{m}$: $AIC(I)=-69.4$, $AIC(II)=-83.7$; $5.8\mu\mathrm{m}$: $AIC(I)=-69.5$, $AIC(II)=-84.5$; and $8.0\mu\mathrm{m}$: $AIC(I)=-72.6$, $AIC(II)=-84.6$. The $SBC$ show similar results as $AIC$. Hence, statistical tests on the possible metallicity dependency of theoretical IRAC band P-L slopes suggested that this dependency could be linear.

\section{Discussion and Conclusion}

The statistical tests suggested that the synthetic IRAC band P-L slopes are linearly depending on $12 + \log(O/H)$. However, this result contradicts to the expectation that the MIR P-L relation should be independent of metallicity. Empirical IRAC band PL slopes, based on the {\it Spitzer} archival data, indicated that these P-L slopes may be metallicity independent \cite{nge10}. Further works, both theoretically and empirically, need to verify or falsify the dependency of metallicity on IRAC band PL relations.

\ack
CCN thanks the funding from National Science Council (of Taiwan) under the contract NSC 98-2112-M-008-013-MY3.

\end{document}